\documentclass{aastex631}

\received{\today}
\submitjournal{ApJ}

\begin{document}

\title{The linear-mixing approximation in silica-water mixtures at planetary conditions}  

\author{Valiantsin Darafeyeu}
\affiliation{Dept. of Astrophysics, University of Zurich, Winterthurerstrasse 190, CH-8057 Zurich, Switzerland}
\affiliation{Institute of Molecular and Clinical Ophthalmology Basel
Mittlere Strasse 91, CH-4031 Basel, Switzerland}

\author{Stephanie Rimle}
\affiliation{Dept. of Astrophysics, University of Zurich, Winterthurerstrasse 190, CH-8057 Zurich, Switzerland}

\author{Guglielmo Mazzola}
\affiliation{Dept. of Astrophysics, University of Zurich, Winterthurerstrasse 190, CH-8057 Zurich, Switzerland}

\author{Ravit Helled}
\affiliation{Dept. of Astrophysics, University of Zurich, Winterthurerstrasse 190, CH-8057 Zurich, Switzerland}



\begin{abstract}

The Linear Mixing Approximation (LMA) is often used in planetary models for calculating the equations of state (EoSs) of mixtures. A commonly assumed planetary composition is a mixture of rock and water. Here we assess the accuracy of the LMA for pressure-temperature conditions relevant to the interiors of Uranus and Neptune. 
We perform MD simulations using {\it ab-initio} simulations and consider pure-water, pure-silica, and 1:1 and 1:4 silica-water molecular fractions at temperature of 3000 K and pressures between 30 and 600 GPa.  
We find that the LMA is valid within a few percent  ($<\sim$5\%) \textcolor{black}{between $\sim$150-600 Gpa,} where the sign of the difference in inferred density depends on the specific composition of the mixture. We also show that the presence of rocks delays the transition to superionic water by $\sim$ 70 GPa for the 1:4 silica-water mixture.  Finally, we note that the choice of electronic theory (functionals) affect the EoS and introduces an uncertainty in of the order of 10\% in density. 
Our study demonstrates the complexity of phase diagrams in planetary conditions and the need for a better understanding of rock-water mixtures and their effect on the inferred planetary composition. 

\end{abstract}

\keywords{materials science, phase transitions, Uranus, Neptune, water, silica, rocks}


\section{Introduction} 

One of the key objectives in planetary science is to determine the bulk compositions and internal structures of planets. 
Planets are typically characterized using planetary models that solve the standard structure equations \cite[see e.g.,][] {Stevenson1982,Guillot2015} together with the equation of state (EoS) of the assumed materials. Since planetary interiors cover a large range of pressures and temperatures and can consist of various elements and their mixtures, for simplicity, the EoS of the mixtures is calculated using the Linear Mixing Approximation (LMA). 

Although the bulk compositions and internal structures of Uranus and Neptune are not well determined, it is clear that their compositions  are not dominated by hydrogen-helium \textcolor{black}{(in mass fraction, not in particle number) but are mostly composed of},  heavier elements such as ices (water, ammonia and methane) and rocks (SiO$_2$, MgSiO$_3$). 
The global water-to-rock-ratio in the planets 
 has a huge range and is very model-dependent. As a result, it is still unclear whether Uranus and Neptune are water-dominated (``icy") or rock-dominated (``rocky"). 
 \textcolor{black}{In addition, it is still unclear whether rocks and water are mixed in the planetary deep interior or are separated into distinct regions. Realistically, it is hard to imagine that a pure-water layer exists. Formation models clearly show that materials are expected to be mixed in planetary interiors and that composition gradients form \citep[][]{2017ApJ...840L...4H,2022ApJ...931...21V}. The behavior of a rock and water mixture depends on the miscibility of the two materials \citep{Vazan2022,kovacevic_miscibility_2022}. At the moment, due to the large range of possible temperature profiles of Uranus and Neptune it remains unclear whether rocks and water will be mixed \citep[e.g.,][]{2019MNRAS.487.2653P, Neuenschwander2024}. Since an interior with rock-water mixtures is possible, confirming the validity of the LMA for rock and water mixtures is desirable}. 

In addition, the EoS tables used for planetary models are typically derived from {\it ab-initio} electronic structure simulations.  However, these simulations are not exact and the inferred EoS depends on the specific approximation (density functional theory, DFT). It is therefore also critical to assess the uncertainty in the EoS linked to the choice of DFT level of theory. 

In this paper, we perform molecular dynamics (MD) simulations of water, rocks, and their mixtures. The rocks are represented by silica SiO$_2$. The simulations include pure-water, pure-rock, and 1:1 and 1:4 silica-water mixtures at a temperature of 3000 K and a pressure range between 30 GPa and 600 GPa.  \textcolor{black}{We also explore the validity of the LMA and their dependency on the assumed  electronic theory, beyond the most commonly used PBE functional. 
The selected range of pressures and temperatures}  corresponds to the deep interiors of Uranus and Neptune and represents water-dominated and rock-dominated rock-water mixtures.  
Note that these configurations are also relevant for modeling the interiors of 
 intermediate-mass exoplanets with rock-water compositions.

\begin{figure}[h!]
    \centering
    \includegraphics[width=0.7\columnwidth]{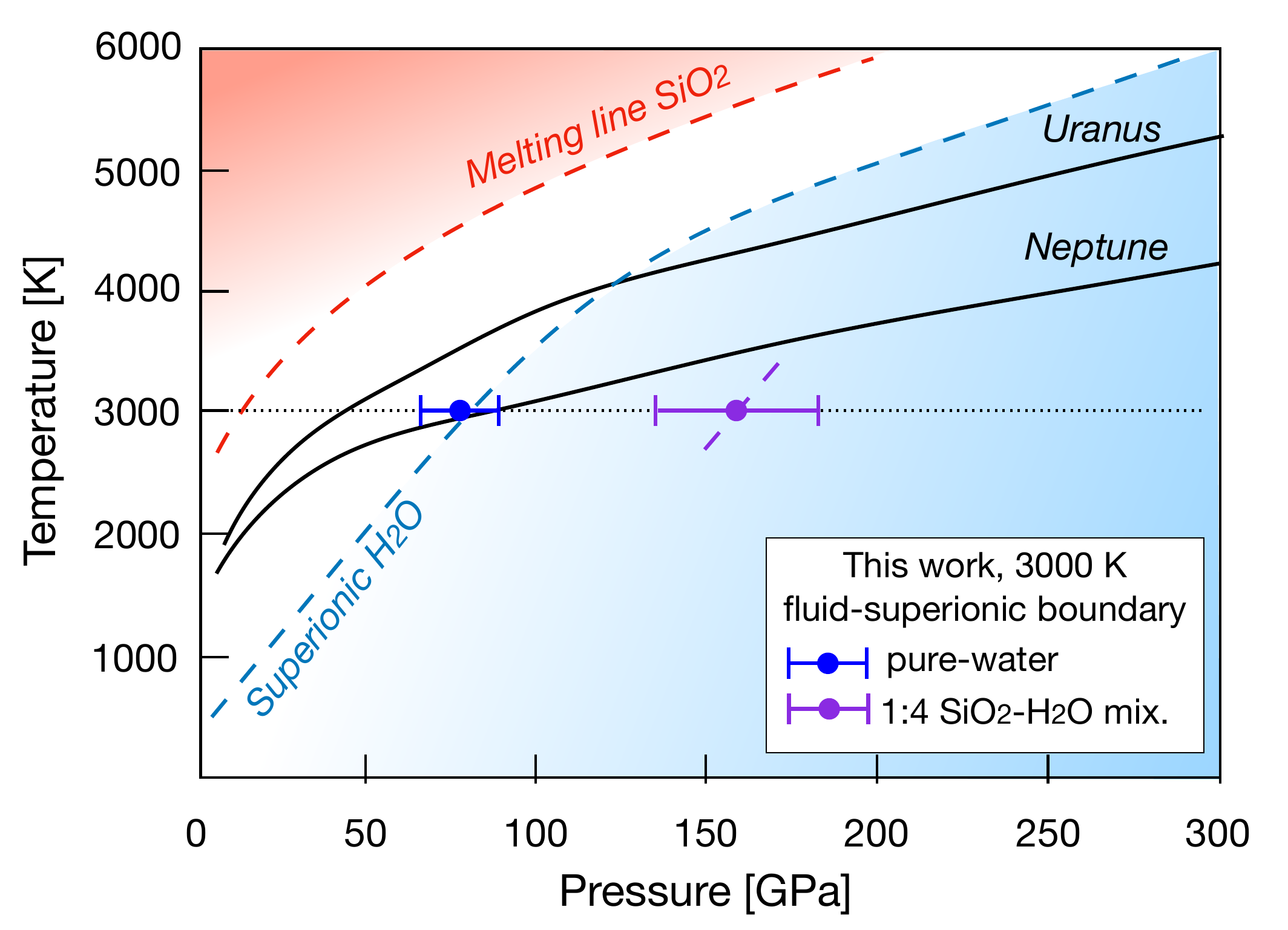}
    \caption{Pressure-Temperature phase diagram showing  Uranus and Neptune adiabats, the fluid-to-superionic transition of pure water (dashed blue) reported in Ref.~~\citep{millot_experimental_2018}, and the melting line of pure SiO$_2$ (dashed red) from Refs.~~\citep{millot2015shock,gonzalez2016melting}.
Simulation performed in this work have been conducted at 3000 K (dotted line). We can observe two phase transitions, the one of pure-water, (blue) which is compatible previous simulations, and another in the 1:4 mixture (purple). The presence of 25$\%$ SiO$_2$ molecules delays the superionic water phase by about 70 GPa. The corresponding phase boundary line is sketched (dashed purple) in the \textcolor{black}{vicinity of 3000 K, assuming a temperature dependence similar to the pure-water case (dashed blue)}. Error bars denote the leftmost (rightmost) point in the pressure line where we observe a fluid (superionic) MD simulation, \textcolor{black}{see Figure}.~\ref{fig:md1}.}
    \label{fig:phase}
\end{figure}

 Materials at GPa pressures exhibit unexpected phase transitions. The most notable discovery, obtained through numerical simulations, is that of the superionic phase of water (see Fig.~\ref{fig:phase}) and  where oxygen occupies crystal positions while hydrogen atoms are free to move within it~\citep{Cavazzoni44}.  This phase is relatively easy to observe in simulations equipped with electronic structure solvers such as DFT~\citep{PhysRevE.93.022140,wilson_superionic_2013,PhysRevLett.117.135503} and was also recently observed  experimentally~\citep{millot2019nanosecond,prakapenka_polymorphism_2020}. 
 The existence of the super-ionic phase in water has significant applications for planetary modeling. 
 The presence of freely moving protons affects the electrical conductivity and therefore the source of the magnetic fields of Uranus and Neptune. 
 \textcolor{black}{It remains unknown whether the magnetic fields of Uranus and Neptune are generated due to ionic water or other elements/mixtures that become electrically conducting in the planetary deep interior. It is of high interest to explore the electrical conductivity of water-rock mixtures as well as refractory elements that are mixed with hydrogen (and helium). }

Since the fluid-superionic transition is driven by electronic structure changes, it is likely that the inclusion of other elements can change the phase diagrams of the pure materials. 
The superionic phase of water was found to be stable when doped with other molecules, such as ammonia~\citep{doi:10.1021/acs.jpca.5b07854}, although this can be expected since also pure-ammonia develops its own superionic phase~\citep{Cavazzoni44,doi:10.1063/1.4810883}. 

Numerical structure-search methods can identify possible ternary or quaternary combinations of elements such as hydrogen, oxygen, carbon, nitrogen which produce chemically stable materials at high pressures and low temperatures~\citep{conway2021rules,naumova2021unusual}. Also, recent experiments revealed the rich physics and chemistry of \textcolor{black}{these materials at conditions relevant for icy giants}~\citep{chau2011chemical}. 

Recently, structure-search simulations have shown that at pressures of $\sim$400-600 GPa, also some ternary structures of silica, oxygen, and hydrogen, with a given ratio can become stable, with the corresponding phase being  superionic~\citep{PhysRevLett.128.035702}.

More simulations are needed to assess the stability of the silica-water superionic phase at lower pressures. In particular, it is important to investigate how the presence of an arbitrary amount of silica in water affects the fluid-to-superionic transition, and whether such a mixed silica-water superionic phase exists at these pressures, as we perform in this study.  

There is also an additional complication. Water and rocks can either mix or separate depending on the temperature/pressure conditions. Recent experimental studies using SiO$_2$~\citep{nisr2020large} and MgO~\citep{kim2021atomic} indicate that these compounds are indeed miscible with water at GPa pressures and intermediate temperatures (below $\sim$ 2000 K). 
However, it is not possible to draw conclusions about the deep interiors of Uranus and Neptune from these studies since they either focus on temperatures~\citep{nisr2020large} or  pressures~\citep{kim2021atomic} that are lower than those expected in the planetary deep interiors. If materials are immiscible at given pressures and temperatures it is expected that this immiscibility would also occur at higher  temperatures, and therefore, possibly in the deep interiors of Uranus and Neptune. However, the behaviour of phase diagrams is complex and it would be important to confirm this theoretically and experimentally. If no phase separation between these elements occur, then their mixture is expected to form a new type of solid phase or, possibly, superionic phase. 

The existence of first-order phase transitions occurring at different pressures in pure-water, pure-silica~\citep{martovnak2006crystal,PhysRevLett.126.035701}, and their mixtures implies that the LMA  must be violated in these regions. The objective of this study is to determine the extent of this error quantitatively. 

\section{Methods}

The simulation of materials at GPa pressure requires an ab-initio treatment of the electronic structure, as electronic-driven phase transitions are expected.
In this work, we treat the nuclei as classical particles (which is a reasonable approximation at temperatures on the order of thousands of Kelvin). The electronic structure is obtained through density functional theory (DFT) calculations, a method that provides a good balance between accuracy and computational cost. 
Furthermore, DFT simulations for water and its superionic phase transition are in reasonable agreement with recent experiments~\citep{millot_nanosecond_2019}.
However, it should be noted that an ab-initio simulation is not synonymous with``exact," as DFT equipped with standard functionals remains an approximate method~\citep{burke_perspective_2012}.
The objective of this work is to explore the validity of the LMA and its dependency on  the level of electronic theory used.
Until now, this aspect \textcolor{black}{has not received}  much attention despite its importance and potential influence on the results. All the previous studies adopt the Perdew-Burke-Ernzerhof (PBE) functional and have not compared to any other theory ~\citep{Cavazzoni44,wilson_superionic_2013,french2016ab, PhysRevLett.117.135503, PhysRevLett.110.151102, PhysRevE.93.022140}. The choice of PBE is reasonable because it is compatible with shock-wave experiments on compressed fluid water~\citep{PhysRevLett.108.091102}. However, it still needs validation over the broader phase diagram and in the presence of mixtures.
For example, it is known that the PBE functional does not reproduce experimental values for water at room temperature~\citep{zen2015ab,morales2014quantum} and underestimates band gaps at high pressure, favoring delocalized electronic states~\citep{morales_nuclear_2013}.
Furthermore, PBE fails to capture the low quartz to coesite transition in high-pressure pure-silica, such that it is expected that electronic quantum Monte Carlo methods can finally reproduce the experimental data~\citep{driver2010quantum}.
Overall, values of electronic pressure calculated with different functionals may vary within a range of about 10$\%$. This potential systematic error is rather large and is comparable (or larger) than the expected accuracy of the LMA.

The molecular dynamics (MD) simulations were performed using cubic cells containing 54 molecules, \textcolor{black}{similarly to previous works assessing the LMA on mixtures of water, ammonia, and methane}~\citep{bethkenhagen_planetary_2017}.
We consider pure-water, i.e., 54 water molecules, pure-silica, i.e., 54 SiO$_2$ molecules and 1:1 and 1:4 silica-water molecular fractions. \textcolor{black}{Note that these ratios correspond to particle numbers and not mass fractions.}
\textcolor{black}{For each compound we start the isotherms at low density, then we decrease the cell's size. The starting atomic configuration at a new density is taken from an equilibrated configuration at the nearest lower density.}
  We perform 10 simulations for each of the four compounds considered, at different densities.
 In order to explore a range of pressures from 30 to 600 GPa, we tune the range of density of the four materials accordingly. 
We first perform the pure-water simulations.
  In order to reduce the equilibration times of the mixture simulations we use the following strategy:  We 
  initialize low-density configurations with 11 SiO$_2$, corresponding to 1:4 molecular and 46\% mass silica-water fraction, and with 27 SiO$_2$, corresponding to 1:1 molecular and 77\% mass silica-water fraction by replacing H$_2$O molecules with SiO$_2$  and performing structural relaxations.
We run the MD in the NVT ensemble, for at least 1-2 ps after the equilibration to measure the pressure and the diffusion coefficients. \textcolor{black}{Longer simulation times of up} to $\sim$8 ps, are used when thermalization is more elusive.
\textcolor{black}{In Appendix~\ref{app:equilibration} we show example of thermalization of our MDs. }
 We adopt the Berendsen thermostat to maintain a temperature of 3000 K, which is relevant for the study of Uranus and Neptune.
 We use time steps in the range 0.5-2.0 fs.
 Interestingly, we observe that the choice of the thermostat is crucial. 
 We use the open-source code Quantum Espresso for our DFT simulations~\citep{giannozzi_quantum_2009}. Gamma point and a cutoff energy of 950 eV are used in all our runs.
 We adopt projector augmented wave pseudopotential for all functionals considered here with the exception of SCAN, where we use optimized norm-conserving Vanderbilt ones~\citep{giannozzi_quantum_2009}.
 The pure-water EoS obtained is perfectly compatible with previous studies~\citep{PhysRevE.93.022140} which employs similar electronic set-ups and a different DFT software.
\textcolor{black}{ In Appendix~\ref{app:test}, we test the convergence of the DFT setup. We use $1 \times1\times1$ and $2\times2\times2$ shifted k-point mesh, for pure water, 1:4 mixture, and pure silica, at pressure of $\sim$200 GPa.  
We also employ a larger energy cutoff using a smaller timestep showing that the simulations presented in the main text are converged.}



\section{The water-silica mixture phase diagram}
In this section, we explore how the transition to superionic water is affected by the existence of \textcolor{black}{silica}. 
\textcolor{black}{Following previous literature, it is expected that materials undergo phase transitions upon compression. The primary objective is to reproduce the well-known fluid-superionic phase transition for pure water and discover if there is qualitatively a phase transition for mixtures as well. Clearly, this investigation is important to test the LMA, which is necessarily violated close to the phase transition. Near the transition, the EoS exhibits discontinuities, and as we will see, these occur at different pressures depending on the type of mixture. Therefore, we expect the LMA to be more accurate in a pressure region away from phase transitions.}

\begin{figure*}[t]
    \centering
    \includegraphics[width=0.85\textwidth]{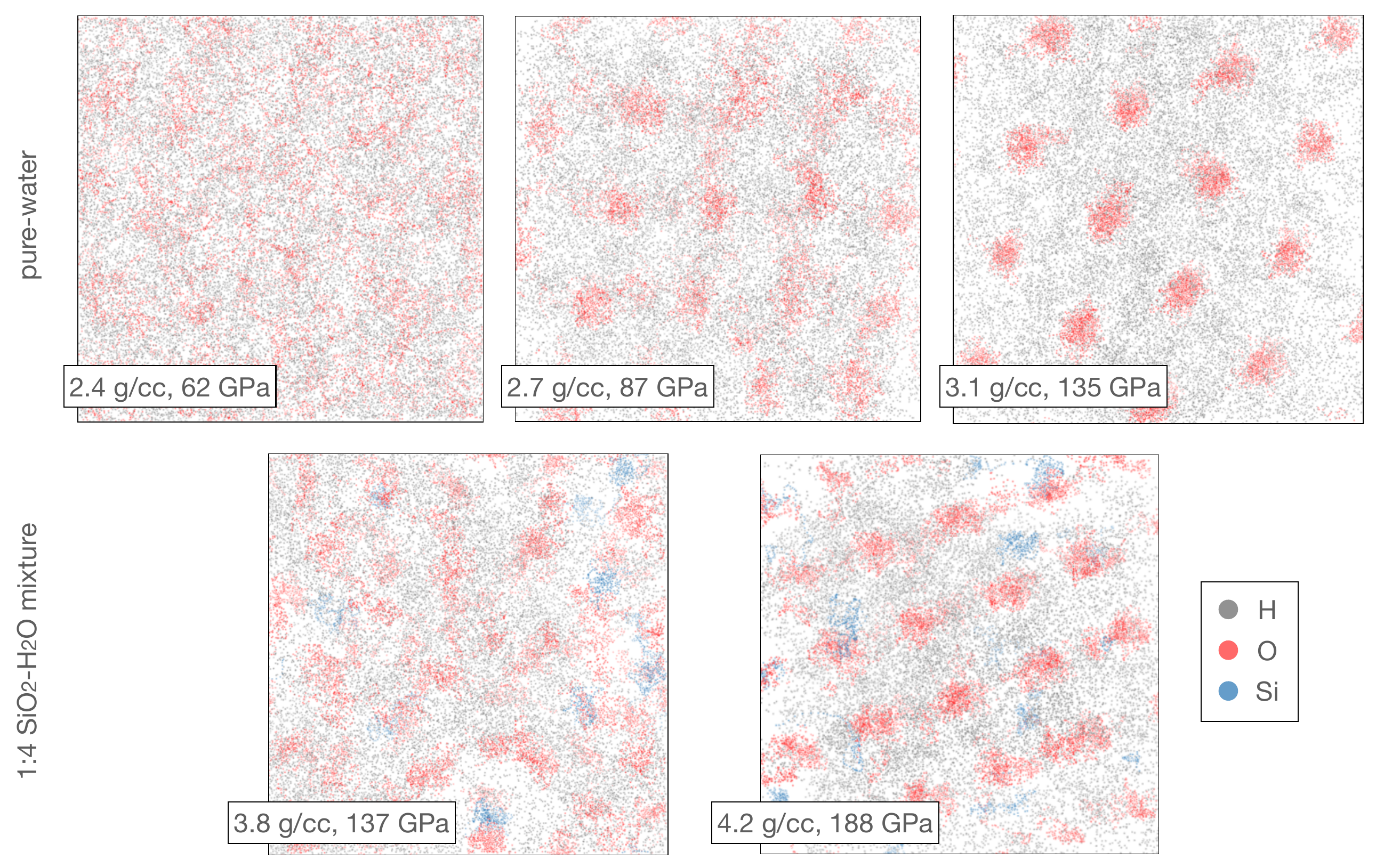}
    \caption{ Positions of the atoms observed during a 500 fs time-window after equilibration has been attained at different densities. Top: pure-water simulations at a fluid (left) and superionic phase (right). \textcolor{black}{The spontaneous formation of the Oxigen lattice is observed at about 87 GPa. We can therefore conclude that the onset of the superionic phase is between 62 and 87 GPa.} Bottom: 1:4 mixture simulations. These observations are used to draw the phase diagram of Fig.~\ref{fig:phase}. Gray, red and blue dots correspond to H, O, and Si atoms respectively.  }
\label{fig:md1}
\end{figure*}

\textcolor{black}{In this work, we primarily rely on trajectory analysis to establish the phase transitions. We also track possible jumps in the oxygen diffusion coefficients, indicating the transition from a fluid to a crystalline structure (for the oxygen atoms).} 
\textcolor{black}{However, in our case, this method of observing phase transitions is less conclusive than a simpler inspection of the structures that spontaneously form during the simulation.}
Analyzing the trajectories it is possible to observe the onset of the superionic oxygen lattice already at 87 GPa, such that the fluid-to-superionic transition occurs between 62 and 87 GPa at 3000 K.
\textcolor{black}{Fig.~\ref{fig:md1} shows a collection of snapshots from the MD simulations at different pressures. It is clear from the figure that a structural transition occurs} where the oxygen atoms are creating a lattice structure  while the hydrogen atoms are freely moving.

The transition of pure-water from liquid to superionic water is also characterized by a significant decrease, by nearly two orders of magnitude,  of the \textcolor{black}{oxygen} diffusion coefficient (see  Appendix, Fig.\ref{fig:diff} (blue curve)).
While we notice that the diffusion coefficient jump is not very abrupt in our simulations and that the finite size of the simulation cell prevents us from making a quantitative statement about the crystal lattice that has been spontaneously formed, our finding is in very good agreement with previous experimental~\citep{prakapenka2021structure} numerical results at the PBE level~\citep{wilson_superionic_2013,french2016ab}. Precisely resolving the type of oxygen's crystal structure~\citep{wilson_superionic_2013,french2016ab,prakapenka2021structure} is outside the scope of this work, given that the free energy difference between competing crystal lattices is negligible compared to the error of the LMA \citep{wilson_superionic_2013}.



After reproducing the pure-water transition we analyze the 1:4 and 1:1 mixture simulations. 
Pure-silica is always in a solid phase at 3000 K and at all pressures considered here~\citep{millot2015shock,gonzalez2016melting,schoelmerich2020evidence}, while pure-water, as mentioned above, undergoes a structural transition from fluid to superionic. 
The presence of SiO$_2$ molecules in water is expected to produce two contrasting tendencies. In small quantities, the presence of SiO$_2$ disrupts the onset of the oxygen lattice and therefore delays the phase transition, stabilizing the liquid at higher pressures.
On the other hand, when the fraction of SiO$_2$ is larger, being below its melting point, the mixture becomes predominantly solid \textcolor{black}{(see Fig.~\ref{fig:md1})}.

Since our method of initializing the mixtures starts with fully mixed H$_2$O-SiO$_2$ molecular structures, we do not expect to observe phase separation phenomena that require time scales longer than the simulation time. 
Experiments show that water and silica are miscible at a temperature of about 2000 K and pressure range of above 20 GPa,
~\citep{nisr2020large,Vazan2022} thus justifying our choice of starting from fully mixed atomic structures.
It is however possible to observe structural rearrangements, which nonetheless indicate the complexity of the chemical reactions between these two molecules as we discuss below.

Tracking the putative phase transition by monitoring the diffusion coefficient of oxygen in that case is more complex than in the case of pure-water.
In the case of the 1:4 mixture, the diffusivity remains stable until $\sim$137 GPa and then decreases from the simulation corresponding to 188 GPa onwards \textcolor{black}{(see Appendix, Fig~\ref{fig:diff})}.
Visual inspection of the sampled atomic structures at the two pressures reveals a spontaneous structural transition from a fluid configuration where oxygen atoms move more freely to one where the oxygen atoms appear in crystalline positions.
Therefore, we can conclude that the presence of SiO$_2$ impurities delays the transition by approximately 70 GPa, compared to the pure-water case.
The diffusion coefficient of hydrogen decreases by a factor of two compared to the fluid phase, in contrast to the behavior of pure-water, where it remains constant during the transition, although it has a comparable order of magnitude \textcolor{black}{(see Appendix, Fig~\ref{fig:diff}, right panel)}.

\begin{figure*}[t]
    \centering
    \includegraphics[width=0.9\textwidth]{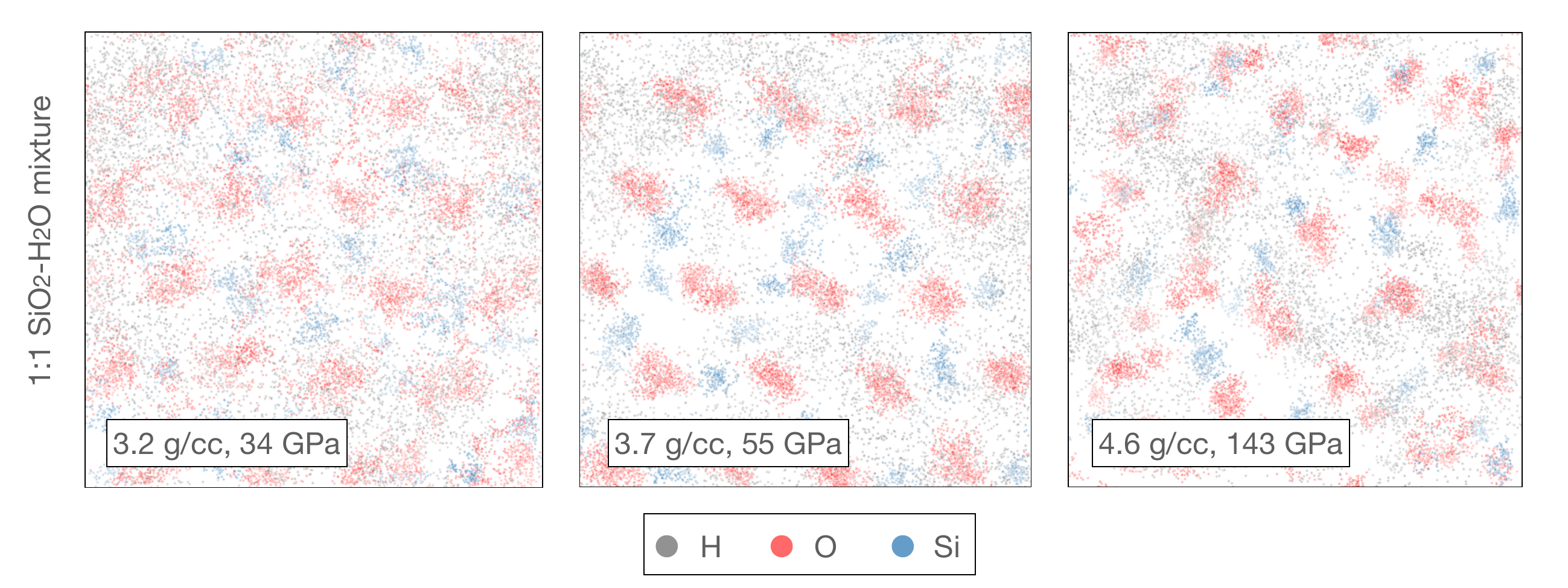}
    \caption{ Positions of the atoms observed during a 500 fs time-window after equilibration has been attained. Here we plot three density conditions for the 1:1 mixture. Gray, red, and blue dots correspond to H, O, and Si atoms respectively. It is possible to observe the increased localization of O atoms, from the 34 to the 55 GPa MD simulation. Even at the lowest pressure, the O atoms seems to perform local fluctuation around a crystal structure.}
\label{fig:md2}
\end{figure*}

The situation for the 1:1 mixture is more complex. At 34 GPa, \textcolor{black}{i.e. at the lowest density considered here}, it is possible to infer already the existence of a lattice structure, although the oxygen atoms undergo significant displacements around the equilibrium positions.
Following compression, at 55 GPa, we observe the formation of a more well-defined lattice of oxygen inter-layered with Si atoms. 
This configuration is characterized by a decrease in the diffusion coefficient of oxygen. 
Subsequently, upon further compression (143 GPa), the oxygen atoms appear to localize again but in a disordered solid or solid with defects.
The diffusion coefficient of oxygen is approximately an order of magnitude higher than that of the superionic water phase, while the one of hydrogen remains similar.
Nevertheless, some volumes appear to be hydrogen-free.
It is therefore not straightforward to accurately assign the phase of the 1:1 mixture above 140 GPa. 
\textcolor{black}{Given the possibility that the small simulation cell introduces frustrations in the lattice, our predictions should be taken as qualitative because the exact transition pressure could change when considering larger cells.}
\textcolor{black}{In addition, further investigations are required to conclusively determine whether the observed high-pressures mixture phases are superionic phases.}

{\color{black}
In Appendix~\ref{app:gr} we present the radial pair distribution functions, $g(r)$, i.e. the thermal distribution of O-O and Si-Si distances, at all density considered. 
Interestingly, the position of the first peak of the O-O $g(r)$ is similar in all mixtures, at same pressure. At low pressures, this observation is consistent with the crystal structure of superionic ice and stishovite, i.e., the phase of SiO$_2$ which is stable at tens of GPa \citep{ross1990high}.
Overall, these radial pair distribution functions suggest that the 1:4 mixture is more similar to pure-water, while the 1:1 mixture is closer to the pure-silica parent compound.
}

\section{The linear mixing approximation}

The LMA for the density of a mixture can be defined as: 
\begin{equation}
\label{eq:lma}
\frac{1}{{\rho}_{\textrm{mix}}^{\textrm{LMA}}(P,T)}=\sum_{i=1}^n\frac{x_i}{{\rho}_{i}(P,T)}, 
\end{equation}
where, $n=2$ for our binary mixture conditions, $x_i$,$u_i$ and ${\rho}_i$ are the mass fraction, specific internal energy, and density of the compound $i$ at given conditions, respectively. 

 To test the validity of the LMA for water-rock mixtures at planetary conditions, we compute the EoS of the 1:1 \textcolor{black}{and} 1:4 mixtures for an isoterm of 3000 K, and we compare with the predictions given by Eq.~\ref{eq:lma}, using the pure-water and pure-silica data. 
 The data points coming from the MD simulations are shown in Fig.~\ref{fig:eos}, and presented in tabular format in Appendix~\ref{app:fulltable}.
 For the pure materials (pure-water, pure-silica) we also plot a cubic fit to data, $P_{\textrm{water}}(\rho)$ and $P_{\textrm{silica}}(\rho)$ .
 Also shown are the water-rock mixtures in the proportion of 1:1 and 1:4 using the LMA, $P_{\textrm{mix}}(\rho)$, obtained by inverting Eq.~\ref{eq:lma}. The figure suggests that the LMA is in excellent agreement with the direct MD data points. 

\begin{figure}[h!]
    \centerline{\includegraphics{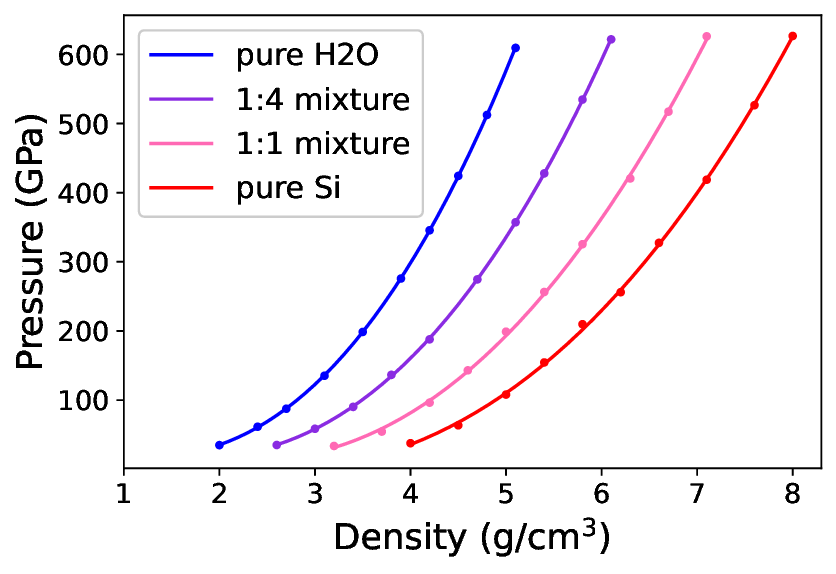}}
    \caption{ 3000 K isotherms for pure water, pure silica and 1:1 and 1:4 mixtures.  The points represent MD data. \textcolor{black}{Statistical error-bars are smaller than the size of the symbol}. For pure-water and pure-silica, the curves correspond to a cubic fit to the data.  For the 1:1 and 1:4 mixtures the lines represent the calculated EoS from LMA. \textcolor{black}{This plot shows how the LMA EoS, for the 1:1 and 1:4 mixtures, is very compatible with the direct mixture MDs. }}
\label{fig:eos}
\end{figure}

\begin{figure}[h!]
    \centerline{\includegraphics{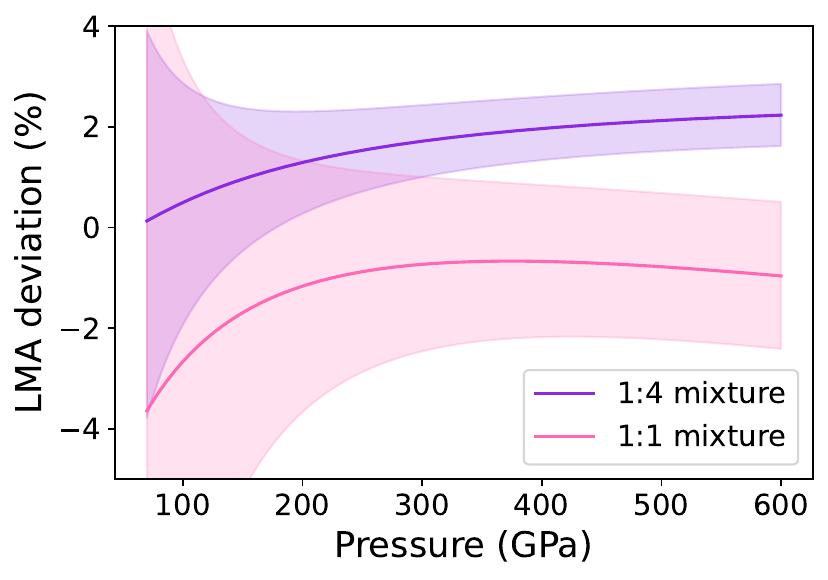}}
    \caption{ Purple and pink curves show the percentage deviation in pressure from linear mixing approximation for 1:4 and 1:1 mixtures respectively. Shading represents the 95\% confidence error band.}
    \label{fig:mix}
\end{figure}

It is important to note that our MD simulations are performed at fixed densities, while pressure is an output. To assess the LMA, we first need to invert the pressure-density relation to obtain a $\rho(P)$ curve for the materials.
The two $\rho_i(P)$ curves for the pure compounds are needed to calculate Eq.~\ref{eq:lma}, while the $\rho_\textrm{mix}(P)$ curves for the mixtures are required in order to validate the prediction.
Since the $\rho(P)$ equations of state are constructed by fitting the data using cubic polynomial functions, the uncertainties in the fitting parameters propagate into a given confidence interval for the LMA.

All quantities needed to estimate the LMA relative error,
\begin{equation}
    {\rho}_{\textrm{mix}}(P) - {\rho}_{\textrm{mix}}^{\textrm{LMA}}(P)\over {\rho}_{\textrm{mix}}^{\textrm{LMA}}(P)
    \label{eq:LMAerr}
\end{equation}
are thus affected by uncertainties coming from the fits. These, in turn, depend on the quality of our MD simulation, although statistical fluctuations are unavoidable.
The pressure error bar for each MD simulation is computed using a binning method and is about 0.2-0.3 GPa.
From the uncertainties of the fit parameters, it is possible to estimate the $95\%$ confidence interval for the fitted function~\citep{chatterjee2013handbook}.
The confidence error bar is then propagated to estimate the difference in Eq.~\ref{eq:LMAerr},
which is plotted as a function of pressure in Fig.~\ref{fig:mix}. 
We find that LMA is satisfied within $3\%$ above 200 GPa. 
At lower pressure, the uncertainty of the estimate increases, due to all the different phase boundaries occurring between 30 and 150 GPa, which makes the computed EoSs less smooth and harder to fit.
Providing such a conservative confidence interval for the validation of the LMA is informative.
For instance, the LMA for the 1:1 mixture is compatible within the error bar with the real fully mixed MD data, at all pressures. The LMA for the 1:4 mixture is, instead, noncompatible with the exact value, but still within a $3\%$ deviation.
\textcolor{black}{Finally, in Appendix~\ref{app:LMA_vs_error} we investigate the robustness of this result against possible, yet unlikely, systematic errors.
These can be introduced by long-lived metastable states, yielding incorrect pressures near phase transitions.}

\section{Choice of DFT functionals} 
In the results presented above, we used the commonly adopted PBE functional. 
In order to assess the importance of the used functionals in MD simulations, we compare the results when using four different possible choices for the exchange-correlation functionals: PBE~\citep{perdew_generalized_1996}, BLYP~\citep{blyp}, VDW-DF2~\citep{vdwdf2} and SCAN~\citep{scan}. 
\par
\textcolor{black}{We perform a new set of MD simulations, using these functionals, and investigate the validity of the LMA at  $\sim$ 420 GPa}


\begin{figure}[h!]
   \includegraphics[height=5.cm]{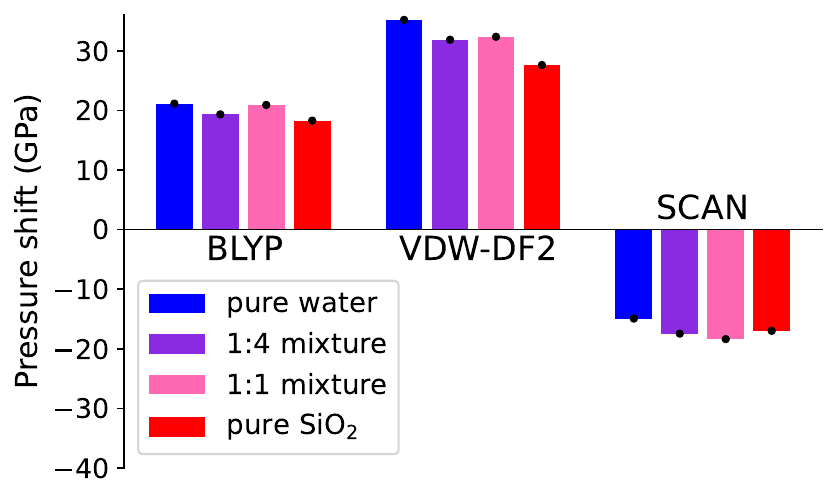}
    \includegraphics[height=5.cm]{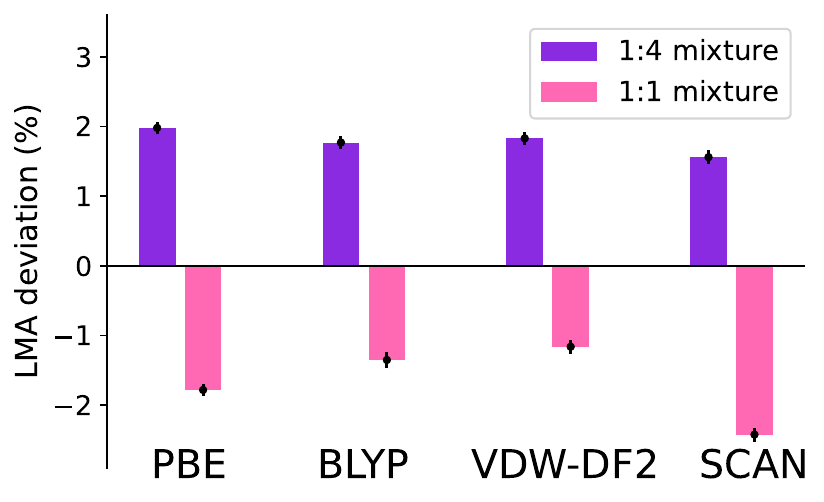}
\caption{{ \textcolor{black}{Left}:} Change of pressure for different functionals compared to PBE pressure (see text). The change in pressure induced by the use of a different functional is similar for all the mixtures. 
{ Right:} Deviation of pressure between the actual pressure data and the LMA for different functionals and the two mixtures. We see that, despite the usage of different functional changes the absolute EoS values, it does not impact as much the validity of the LMA. This occurs because a given functional shifts the calculated pressure of each compound in the same way. }
\label{fig:p_shift}
\end{figure}

The results are shown in Fig.~\ref{fig:p_shift}. 
The top panel shows the pressure shift for the different materials. We find that the change in pressure is comparable in magnitude for a given functional, and is not affected by the amount of SiO$_2$. 
Nevertheless, it is clear that the use of different functionals is significant for the EoSs calculation with a total difference of about 10\%. For example, VDW-DF2 predicts higher pressures at a fixed density, therefore suggesting less dense materials. On the other hand, SCAN always predicts denser materials for the same pressure conditions. 
This effect has already been observed in the study of hydrogen~\citep{Knudson2018,mazzola_phase_2018,morales_nuclear_2013,PhysRevB.89.184106} and hydrogen-helium mixtures~\citep{clay2016benchmarking,PhysRevLett.120.115703}.
This can be understood as a systematic error, rooted in the approximations made at the electronic structure level, in solving the many-electron \textcolor{black}{Schrödinger} equation.
Each exchange-correlation functional implies a slightly modified equation to solve, which produces different electronic densities around the ions, and modified energy and stress operators to be computed.
This results in different values of energy and pressure for the same atomic configuration, and therefore different ensemble values.
DFT predictions are expected to converge in the high-density limit, where the system approaches a non-interacting electron gas.~\citep{mcmahon_properties_2012}
\par

The bottom panel shows the inferred difference in density for these different functionals and the calculated LMA. 
The errors are estimated using the statistical error on pressure computed from the MD simulations. We find that  for all assumed functionals, at 420 GPa, the deviation from LMA is about 2\% (cfn.~Fig.\ref{fig:mix}).
\textcolor{black}{Interestingly, we observe that the SCAN functional leads to the largest deviation from the LMA, for the 1:1 mixing, but the lowest one for the 1:4 mixture.}

\textcolor{black}{Notice that, in this study, we cannot determine which XC-functional is the best for these compounds. The prevalence of PBE as the most common choice does not imply that it is necessarily the best functional to be used.}
Experiments at the moment are not able to benchmark ab-initio EoS over such broad range of pressures, so we expect that validation of DFT EoS should be provided by higher level of theories, such as quantum Monte Carlo~\citep{driver2010quantum,PhysRevB.89.184106,mazzola_phase_2018} \textcolor{black}{which are significantly more computationally expensive.}
\textcolor{black}{However, the current results provide insight into how the LMA is affected by the use of different XC-functionals. We find that the impact of a different XC-functional on the LMA error is smaller compared to the absolute error they introduce to the individual equations of state (EOSs).}

\section{Discussion \& Conclusions} 

Although this study represents a step forward in our understanding of the phase diagram of rock-water mixtures, it is clear that some aspects could be further improved in future studies. 

First, our results are valid under the assumption that our {\it ab-initio} simulations correctly and realistically represent the phases under the explored  thermodynamic conditions.
The setup we used is fairly standard, but nevertheless, the simulation cells, cubic in shape and containing 54 molecules, can certainly introduce frustrations that favor one phase over another. However, the fact that our simulations of pure-water reproduce the fluid-superionic phase boundary from other numerical studies~\citep{redmer2011phase,PhysRevE.93.022140} validates our simulations setup and the inferred results. 
\par 

Second, another important aspect concerns the duration of the simulations. As mentioned earlier, it is unrealistic to expect to observe possible spontaneous\footnote{By "spontaneous" we mean that the sequentially generated ionic configurations in the MD simulations change qualitatively in nature without any prior input.} de-mixing starting from a fully mixed configuration. 
\textcolor{black}{Spontaneous demixing has been numerically observed for hydrogen-helium mixtures and in simulations involving thousands of atoms}\citep{PhysRevB.84.235109}

Based on experimental results, we consider the mixtures as fully mixed and run molecular dynamics (MD) simulations to observe structural phase transitions and compute the EoS. By construction, our numerical experiment can not probe the possibility of silica and water demixing.  Indeed, the system size considered here is too small to observe cluster formations. Nucleation processes that lead to phase separation also occurs on timescale which are several order of magnitudes longer than local molecular rearrangements, i.e. unreachable by picosecond-long simulations. 

A robust way to calculate demixing or melting using direct MD simulations is to compute free-energy or perform hysteresis plots, using sufficiently large simulation  cells~\citep{cheng2020evidence}. This is, however, much more computationally demanding, but could be obtained using surrogate models and machine learning~\citep{cheng2020evidence,deringer2021origins,cheng_phase_2021}.  
We can therefore interpret the possible discrepancy between experiments~\citep{nisr2020large,kim2021atomic} and a recent numerical work~\citep{kovacevic_miscibility_2022} that suggests a miscibility temperature between rocks (MgSiO$_3$) and water above 5000 K. 
As discussed in ~\citet{kovacevic_miscibility_2022} this miscibility temperature should be taken as an upper bound if computed using standard coexistence phase setups.  \textcolor{black}{Therefore, given that in this work we used SiO$_2$ and not MgSiO$_3$ to represent rocks and given that they provide an upper bound for the temperature it is still possible that rocks and water are miscible at 3000 K.} 
Note that in this study we chose a temperature of 3000 K since several interior models suggest that this temperature corresponds to the region in Uranus and Neptune where we expect water and rocks \citep[e.g.,][]{podolak_effect_2019}. However, since non-adiabatic models of the planets suggest that the internal temperatures could be significantly higher, it is desirable that future studies consider higher temperatures \citep[e.g.,][]{Neuenschwander2024}. 
\par 

Finally, note that for the simulations we considered only one temperature of 3000 K. While this already allows us to probe the phase diagram of rock-water mixtures and assess the validity of the LMA, it is clear that a larger range of temperatures (and pressures) should be considered. 
Similarly, this work accounted only for silica and water while in reality, more complex mixtures of elements could exist inside planets. We therefore suggest that future studies should consider more complex mixtures, although it should be noted that for this the total size of the system should increase accordingly. Finally, in this work rocks were represented by SiO$_2$ but other rocky materials such as MgSiO$_3$ should be considered. 
\par

While further studies are required, the following key conclusions can be made: 
\begin{itemize}
    \item The LMA for the cases considered here is valid within a few percent  ($<\sim$5\%), \textcolor{black}{above $\sim 100$ GPa.}
      \item The sign of the difference in inferred density depends on the specific composition of the mixture. 
      \item The presence of some silica \textcolor{black}{(in the 1:4 molecular ratio) stabilizes the fluid phase} and delays the phase transition by $\sim$ 70 GPa at 3000 K. 
      \item The chosen functionals affect the EoS and introduces a systematic  uncertainty of the order of 10\% in density. 
\end{itemize}

This study clearly reveals the complexity of phase diagrams in planetary conditions. 
\textcolor{black}{For instance, the two mixtures (1:4 and 1:1) exhibit different quantitative and qualitative behaviors, both concerning the LMA approximation and, most notably, the phase diagram. We find that the presence of a small (although not negligible) molecular fraction of silica in water stabilizes the fluid phase up to at least 137 GPa. This is counter-intuitive if we consider that pure silica is already solid at the same pressures. However, we can still explain this observation if we regard the silica atoms as impurities that delay the onset of the ordered oxygen crystal structure. This phenomenon is reminiscent of how a solar concentration of helium atoms delays the atomization and metallization transition in compressed liquid hydrogen\citep{PhysRevB.84.235109,mazzola_phase_2018}.}

\textcolor{black}{The 1:1 mixture, on the other hand, displays a crystalline oxygen lattice occurring already at about 30 GPa and is more similar to the parent pure silica material. It is clear, therefore, that the linear mixing approximation holds fairly well for the EoS but is qualitatively incorrect in predicting, for example, phase boundaries, as the two mixtures exhibit opposite behavior regarding the onset of oxygen crystal formation upon compression.}

It is yet to be determined whether Uranus and Neptune are more "icy" or more "rocky" \textcolor{black}{and what materials allow the existence of their magnetic fields} \cite[e.g.,][]{helled_interior_2011,2020RSPTA.37890489T,helled_interiors_2020}. \textcolor{black}{It is therefore of great importance to infer other thermodynamical  and transport properties of rock-water mixtures such as the electrical conductivity, diffusivity, etc.}
Future research can put limits on the maximum rock-to-water ratio in the planets that can still lead to a global magnetic field. This, together with future data from space missions, can be used to further constrain the bulk compositions and internal structures of Uranus and Neptune. 
Finally, improved knowledge of phase diagrams at planetary conditions can be used to guide the assumptions made by planetary modelers and for a better characterization of planets in the solar system and beyond.

\appendix

\section{Diffusion coefficients}

\begin{figure}[h!]
   \includegraphics[height=5.cm]{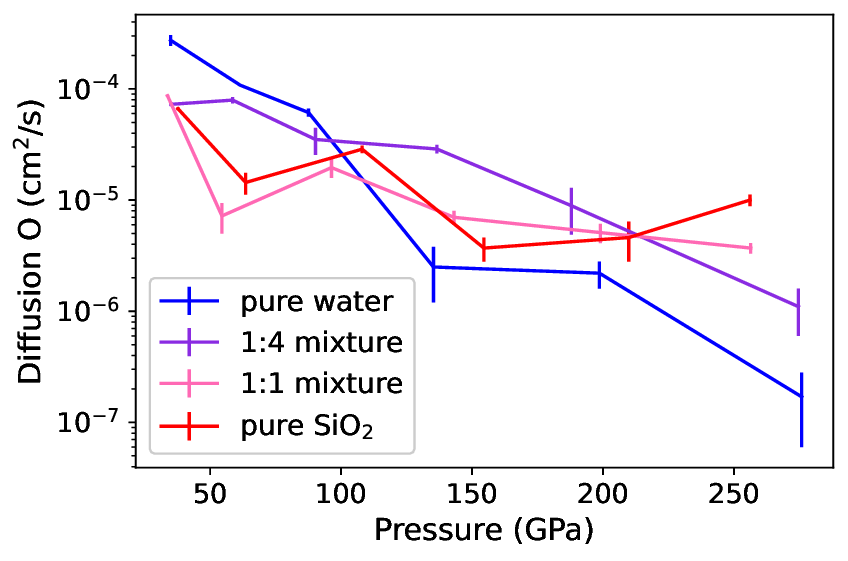}
    \includegraphics[height=5.cm]{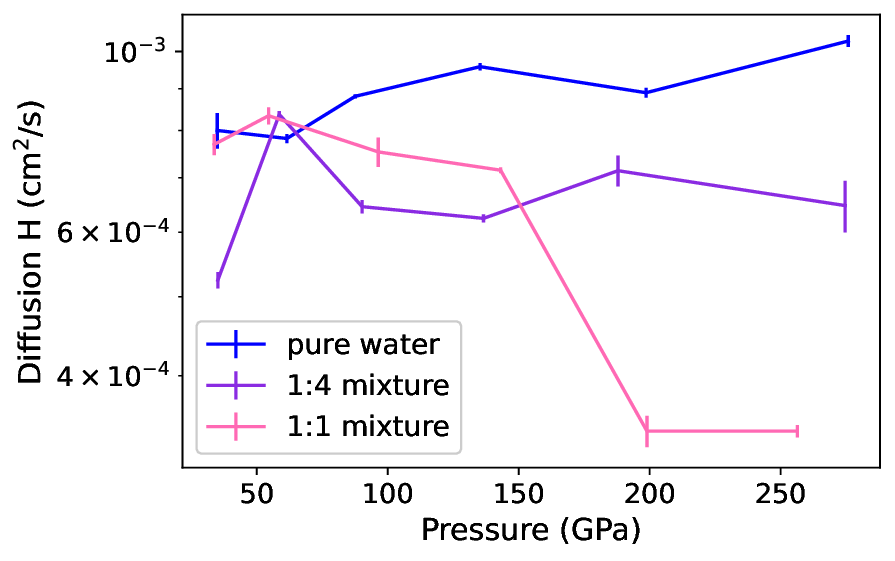}
\caption{{ \textcolor{black}{Left}:} Diffusion coefficients of oxygen for pure-water, pure-silica, and 1:1 and 1:4 mixtures as a function of pressure.
Between 100 and 150 GPa there is a significant drop in diffusion of oxygen following the
fluid-superionic transition. The situation is less clear for the mixture cases.
\textcolor{black}{Right:} Diffusion coefficients of hydrogen for pure-water, 1:1 and 1:4 mixtures as a function of pressure. All the diffusion coefficients are of the same order magnitude within this pressure range. }
\label{fig:diff}
\end{figure}

In Fig.~\ref{fig:diff} we plot the Oxygen and Hydrogen (except for pure-silica) diffusion coefficients as a function of Pressure.

\section{Convergence of DFT calculations}
\label{app:test}
\begin{figure}[h!]
    \includegraphics[height=9.cm]{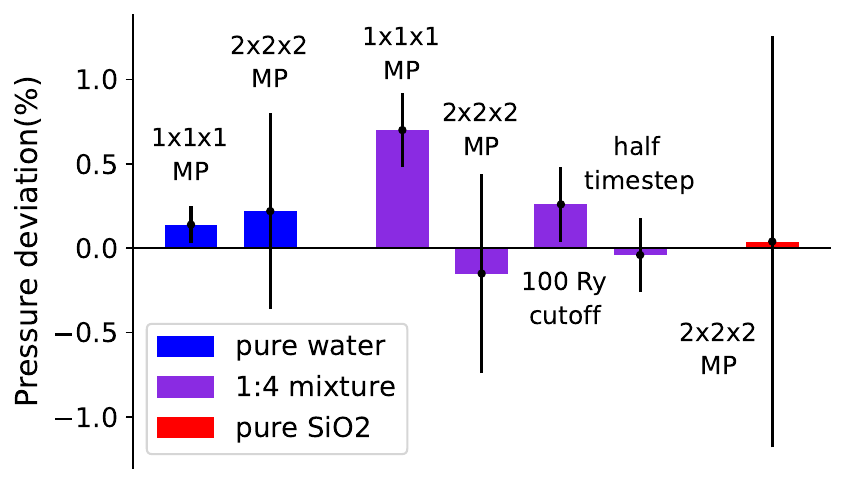}
\caption{Convergence of DFT calculations with respect to the number of k-points, timestep and cutoff. The tests are done for pure SiO$_2$, pure water and 1:1 mixture. MP stands for Monkhorst-Pack k-point grids. The pressure is compared to the baseline pressure at gamma point, 70 Ry cutoff and 0.5 fs timestep (1 fs for 1:1 mixture and 2 fs for pure SiO$_2$).}
\label{fig:dft_test}
\end{figure}

The parameters investigated are the number of k-points, the plane wave cutoff, and the timestep used in the molecular dynamics. In the main text, we adopt the Gamma point for the Brillouin zone sampling. In the absence of exact crystal symmetries, there is no symmetry-related k-point reduction, and any regular grid of points should be equally efficient. We first check that the results do not depend on the position of the k-point by also testing the shifted 1x1x1 configuration. We then increase the number of k-points up to a 2x2x2 grid. The pressure we obtained is consistent with the Gamma point, yielding a value that differs by less than 1$\%$ (and is within error bars) for all three compounds considered.  Second, we investigate the convergence as a function of the used number of plane waves. The DFT simulations we use, Quantum ESPRESSO \citep{giannozzi_quantum_2009}, employs a plane-wave basis set instead of localized ones. Therefore, the number of plane waves is a systematic parameter to check the basis set convergence. The simulation which used a higher number of plane waves, indicated by a cutoff of 100 Ry instead of 70 Ry, also yields compatible results. 
Finally, for the 1:1 mixture, we also perform an MD run with a smaller timestep, i.e., half the original timestep. This ensures that we are not introducing a timestep error. 
Fig.~\ref{fig:dft_test} shows the results of additional MD simulations performed with different DFT and MD set-ups. 

\section{Equilibration of molecular dynamics simulations}
\label{app:equilibration}
\begin{figure}[h!]
    \includegraphics[height=14.cm]{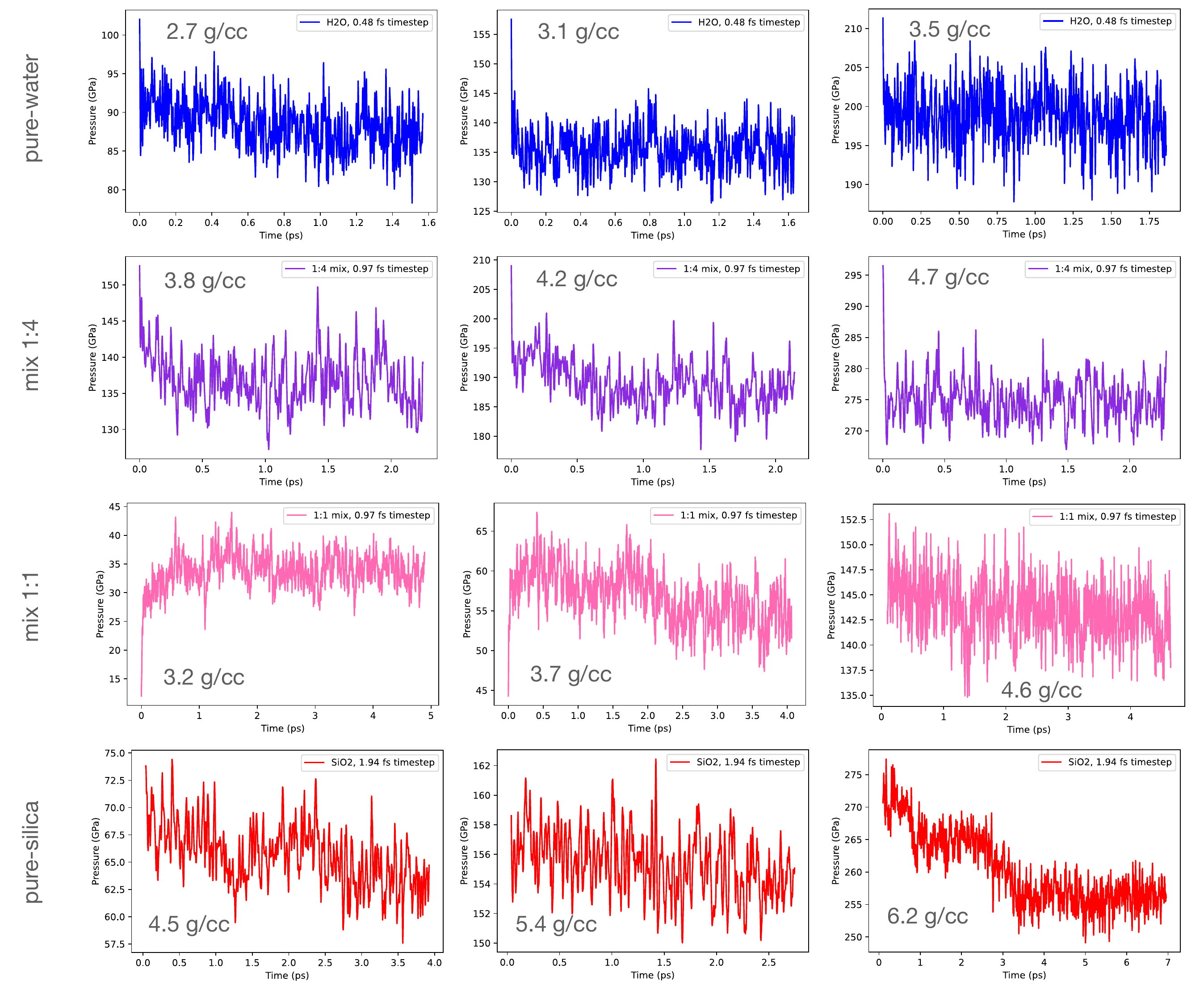}
\caption{Convergence of the MD as a function of simulation time.
The total time of the simulation vary as needed to observe equilibration.
We plot three density conditions for each compound (mostly in the relatively-low pressure regime, where signatures of phase transition occurs) In the case of water and the mixtures the densities here considered are the one featured in Fig.~\ref{fig:md1} and Fig.~\ref{fig:md2}.
The simulation of the 1:1 mixture at $3.7$ g/cc shows a longer equilibration time due to the atoms rearrangement.
The pressure is evaluated after that equilibration has occurred.
}
\label{fig:md_test}
\end{figure}

\begin{figure}[h!]
    \includegraphics[height=12.cm]{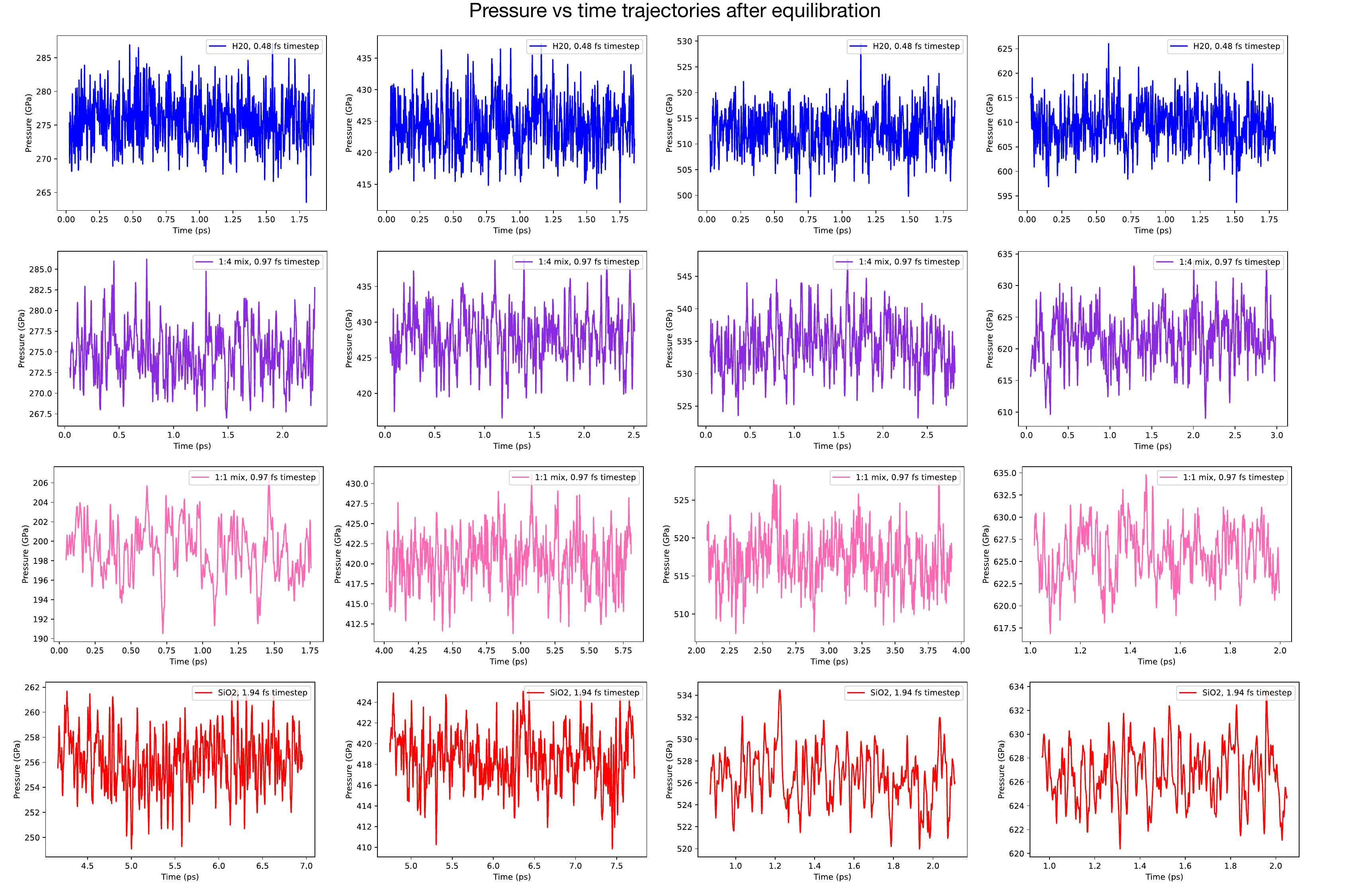}
\caption{ {\color{black} Absence of drifts after the equilibration phase.
Here we plot the pressure vs time, \emph{after} the onset of equilibration. Thermodynamic quantities are evaluated in this time interval. For each compound we plot on simulation at about 250, 420, 530, and 620 GPa (from left to right).}
}
\label{fig:md_equil_proof}
\end{figure}

`
In Fig.~\ref{fig:md_test} we provide an example of the simulations demonstrating a correct equilibration of the MDs. We plot the Pressure as a function of the simulation time. We observe that the pure-water simulations features very quick relaxation times, while the ones with more silica molecules may be trapped in long-lived metastable states.
Each molecular dynamics simulation is initialized using an equilibrated configuration at a lower density, which allows us to reduce the equilibration time. The compression pathways for each mixture are therefore independent of each other. The consistency of the radial pair distribution for different mixtures provides further indirect evidence that no compound has been trapped in a metastable state during compression. Fig.~\ref{fig:md_equil_proof} indicates that the simulations show no sign of drifts after the equilibration regime is reached. This allows us to average the pressure on a time window larger than 2 ps in all cases.

\section{Full results of molecular dynamics in tabular format}
\label{app:fulltable}

In this Appendix we provide the full data in tabular format used to plot Fig.~\ref{fig:eos} and Fig.~\ref{fig:diff}.
Notice that we only report diffusion coefficients below about 300-350 GPa.

\begin{table}[h!]
    \begin{tabular}{|c|c|c|c|c|c|c|}
        \hline
         $\rho$(g/cc) & P(GPa) & error (GPa) & $D_O$(cm\textsuperscript{2}/s) & error (cm\textsuperscript{2}/s) & $D_H$(cm\textsuperscript{2}/s) & error (cm\textsuperscript{2}/s) \\
        \hline
        2.0 & 34.93 & 0.11 & $2.74 \times 10^{-4}$ & $0.30 \times 10^{-4}$ & $8.0 \times 10^{-4}$ & $0.4 \times 10^{-4}$ \\
        2.4 & 61.51 & 0.11 & $1.08 \times 10^{-4}$ & $0.01 \times 10^{-4}$ & $7.82 \times 10^{-4}$ & $0.10 \times 10^{-4}$ \\
        2.7 & 87.48 & 0.15 & $6.13 \times 10^{-5}$ & $0.51 \times 10^{-5}$ & $8.81 \times 10^{-4}$ & $0.05 \times 10^{-4}$ \\
        3.1 & 135.3 & 0.2 & $2.5 \times 10^{-6}$ & $1.3 \times 10^{-6}$ & $9.58 \times 10^{-4}$ & $0.10 \times 10^{-4}$ \\
        3.5 & 198.6 & 0.2 & $2.2 \times 10^{-6}$ & $0.6 \times 10^{-6}$ & $8.90 \times 10^{-4}$ & $0.12 \times 10^{-4}$ \\
        3.9 & 275.8 & 0.2 & $1.7 \times 10^{-7}$ & $1.1 \times 10^{-7}$ & $1.03 \times 10^{-3}$ & $0.17 \times 10^{-4}$ \\
        4.2 & 345.6 & 0.2 & $2.7 \times 10^{-7}$ & $2.0 \times 10^{-7}$ & $9.04 \times 10^{-4}$ & $0.13 \times 10^{-4}$ \\
        4.5 & 424.3 & 0.2 & - & - &- &- \\
        4.8 & 512.4 & 0.2 &- &- &- & -\\
        5.1 & 609.4 & 0.2 &- &- &- & -\\
        \hline
    \end{tabular}
    \caption{Pure water MD simulations with DFT-PBE.}
\end{table}

\begin{table}[htbp]
   \begin{tabular}{|c|c|c|c|c|c|c|}
        \hline
         $\rho$(g/cc) & P(GPa) & error (GPa) & $D_O$(cm\textsuperscript{2}/s) & error (cm\textsuperscript{2}/s) & $D_H$(cm\textsuperscript{2}/s) & error (cm\textsuperscript{2}/s) \\
        \hline
        2.6 & 35.1 & 0.2 & $7.27 \times 10^{-5}$ & $0.28 \times 10^{-5}$ & $5.24 \times 10^{-4}$ & $0.12 \times 10^{-4}$ \\
        3.0 & 58.6 & 0.2 & $7.93 \times 10^{-5}$ & $0.52 \times 10^{-5}$ & $8.37 \times 10^{-4}$ & $0.08 \times 10^{-4}$ \\
        3.4 & 90.2 & 0.3 & $3.51 \times 10^{-5}$ & $0.96 \times 10^{-5}$ & $6.45 \times 10^{-4}$ & $0.12 \times 10^{-4}$ \\
        3.8 & 136.6 & 0.3 & $2.88 \times 10^{-5}$ & $0.25 \times 10^{-5}$ & $6.24 \times 10^{-4}$ & $0.07 \times 10^{-4}$ \\
        4.2 & 187.9 & 0.3 & $8.9 \times 10^{-6}$ & $4.0 \times 10^{-6}$ & $7.1 \times 10^{-4}$ & $0.3 \times 10^{-4}$ \\
        4.7 & 274.6 & 0.2 & $1.1 \times 10^{-6}$ & $0.5 \times 10^{-6}$ & $6.5 \times 10^{-4}$ & $0.5 \times 10^{-4}$ \\
        5.1 & 357.2 & 0.3 &- &- &- &- \\
        5.4 & 427.9 & 0.3 &- &- &- &- \\
        5.8 & 534.6 & 0.3 &- &- &- &- \\
        6.1 & 621.7 & 0.3 &- &- &- &- \\
        \hline
    \end{tabular}
    \caption{1:4 mixture MD simulations with DFT-PBE.}
\end{table}

\begin{table}[htbp]
   \begin{tabular}{|c|c|c|c|c|c|c|}
        \hline
         $\rho$(g/cc) & P(GPa) & error (GPa) & $D_O$(cm\textsuperscript{2}/s) & error (cm\textsuperscript{2}/s) & $D_H$(cm\textsuperscript{2}/s) & error (cm\textsuperscript{2}/s) \\
        \hline
        3.2 & 33.8 & 0.2 & $8.67 \times 10^{-5}$ & $0.16 \times 10^{-5}$ & $7.7 \times 10^{-4}$ & $0.2 \times 10^{-4}$ \\
        3.7 & 54.5 & 0.2 & $0.72 \times 10^{-5}$ & $0.22 \times 10^{-5}$ & $8.3 \times 10^{-4}$ & $0.2 \times 10^{-4}$ \\
        4.2 & 96.4 & 0.3 & $1.96 \times 10^{-5}$ & $0.38 \times 10^{-5}$ & $7.5 \times 10^{-4}$ & $0.3 \times 10^{-4}$ \\
        4.6 & 143.1 & 0.2 & $7.0 \times 10^{-6}$ & $1.0 \times 10^{-6}$ & $7.15 \times 10^{-4}$ & $0.06 \times 10^{-4}$ \\
        5.0 & 199.0 & 0.3 & $5.1 \times 10^{-6}$ & $1.0 \times 10^{-6}$ & $3.42 \times 10^{-4}$ & $0.15 \times 10^{-4}$ \\
        5.4 & 256.4 & 0.3 & $3.7 \times 10^{-6}$ & $0.4 \times 10^{-6}$ & $3.42 \times 10^{-4}$ & $0.06 \times 10^{-4}$ \\
        5.8 & 325.5 & 0.4 &- &- &- &- \\
        6.3 & 420.6 & 0.3&- &- &- &- \\
        6.7 & 517.2 & 0.3&- &- &- &- \\
        7.1 & 626.1 & 0.3 &- &- &- &- \\
        \hline
    \end{tabular}
\caption{1:1 mixture MD simulations with DFT-PBE.}
\end{table}

\begin{table}[htbp]
  \begin{tabular}{|c|c|c|c|c| c | c|}
        \hline
         $\rho$(g/cc) & P(GPa) & error (GPa) & $D_O$(cm\textsuperscript{2}/s) & error (cm\textsuperscript{2}/s) \\
        \hline
        4.0 & 37.8 & 0.2 & $6.6 \times 10^{-5}$ & $0.2 \times 10^{-5}$ \\
        4.5 & 63.6 & 0.2 & $1.4 \times 10^{-5}$ & $0.3 \times 10^{-5}$ \\
        5.0 & 108.0 & 0.3 & $2.9 \times 10^{-5}$ & $0.2 \times 10^{-5}$ \\
        5.4 & 154.5 & 0.2 & $3.7 \times 10^{-6}$ & $0.9 \times 10^{-6}$ \\
        5.8 & 209.8 & 0.3 & $4.6 \times 10^{-6}$ & $1.8 \times 10^{-6}$ \\
        6.2 & 256.1 & 0.2 & $1.0 \times 10^{-5}$ & $1.2 \times 10^{-6}$ \\
        6.6 & 327.4 & 0.3 & - & -\\
        7.1 & 418.7 & 0.3 & - &- \\
        7.6 & 526.5 & 0.3 &- & - \\
        8.0 & 626.6 & 0.2 &- & - \\
        \hline
    \end{tabular}
    \caption{Pure silica MD simulations with DFT-PBE.}
\end{table}

\section{Sensitivity analysis of systematic errors}
\label{app:LMA_vs_error}

\begin{figure}[h!]
    \includegraphics[height=11.cm]{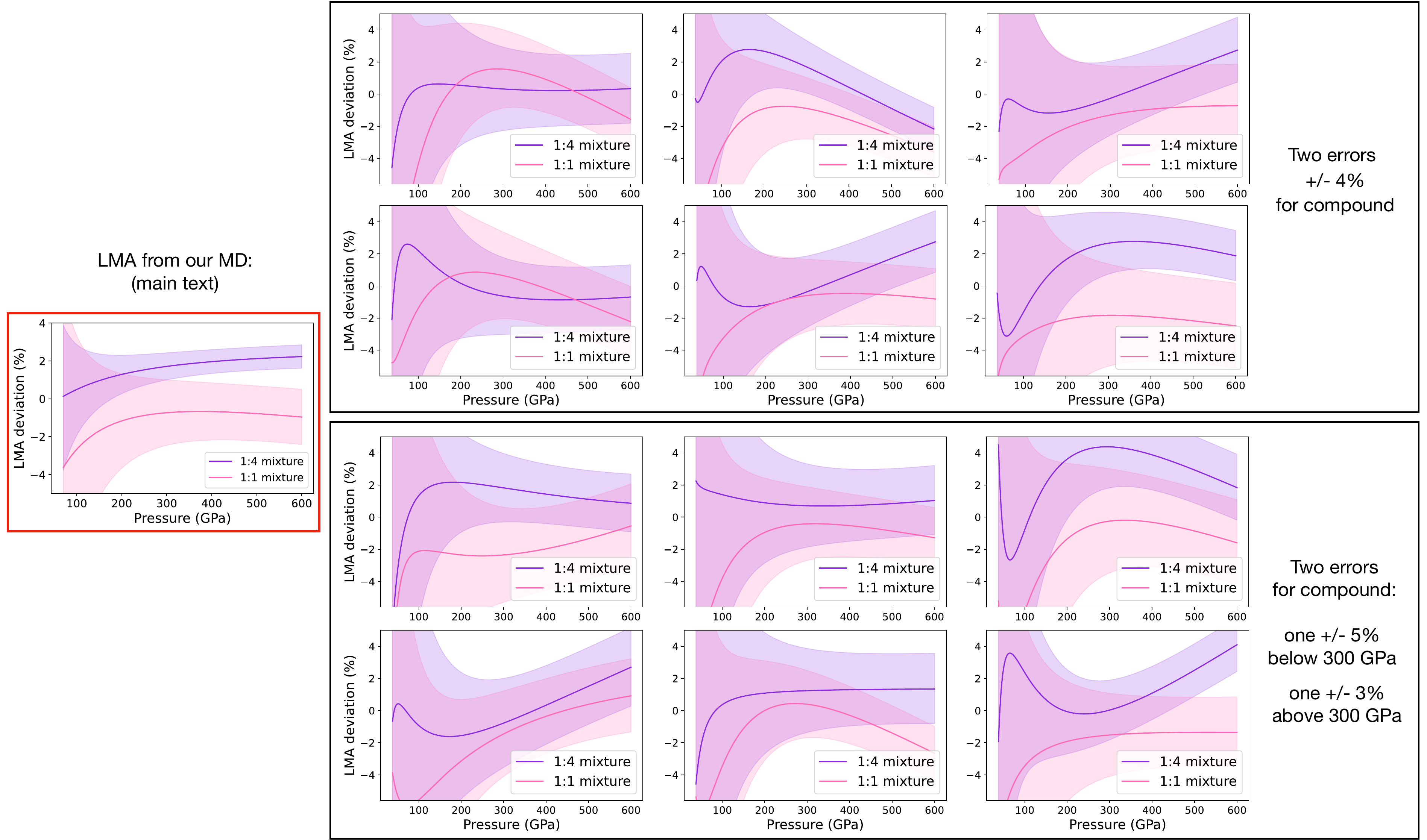}
\caption{Robustness of the LMA against possible systematic errors on estimating some pressures.
Left: Fig.~\ref{fig:mix} from the main text. Right: Scenarios obtained by changing artificially two pressures data points per compound.
}
\label{fig:lma_sensitivity}
\end{figure}

{ \color{black} 
As we have seen, the error on the deviation from the LMA accumulates in a non-trivial way starting from the errors on the individual pressure estimates. The confidence bands we report in the main text figure derive from the statistical errors on the pressure of the simulations.

In this section, we investigate  how  our assessment would vary if we had made two evaluation errors for the pressure of each compound. These could arise from an incorrect equilibration of the MD, for example, in the case of phase transitions that we were unable to explore. 

For the sake of concreteness, if we look at the bottom right trajectory in Fig.~\ref{fig:md_test}, we observe that there is a long-lived metastable phase in the 1.0 to 2.5 ps time window, which is characterized by an average pressure of $\sim$265 GPa. At around 3 ps, the system relaxes toward a phase that yields a pressure of 255 GPa. If we had stopped earlier, or if the metastable phase we encountered had been more resilient, we would have made a non-statistical error of about 4$\%$.

From a conceptual point of view, it can't be guaranteed that all the phases we observe are not  metastable. Even when extending all the simulations for a $10 \times$ total time would not give us this guarantee. 
We can therefore simulate this possibility and investigate the sensitivity of our prediction for the LMA for such potential errors. 
Since we observe a signature of about two phase transitions for each compound, we now assume that two pressures per material are affected by such a systematic error. Therefore, in total, we assume that we have mislabeled 8 pressures in the entire dataset. This is clearly an unlikely scenario, but allows us to validate the robustness of our results. 

In one numerical experiment we extract two $\pm 4 \%$ shifts for all compounds, and we recalculate the EOS fits and the LMA deviation and performed this extractions several times. The results are presented in Fig.~\ref{fig:lma_sensitivity} shows the results for six outcomes. 
In another experiment we extract one  $\pm 5 \%$ shift per compound below 300 GPa and one $\pm 3 \%$ shift per compound above 300 GPa, i.e., we assume that low density simulations are more challenging that the others as far as equilibration times are concerned.
Also for this case we plot six of those outcomes.
We find that these 12 examples are fully consistent with the results we have observed in the main text. 
We can therefore draw two important conclusions:\\

(1) Even introducing as many as 8 mislabelings of about $4\%$, the LMA deviation is always consistent with the results reported above. Therefore, the main result of the paper is robust against mislabeling due to incorrect sampling of the phase diagram and finite time simulations.\\

(2) The plots we obtain after introducing these artificial errors appear much less smooth than the original result, and with much larger confidence bars. This is an indirect evidence for the quality of the original data, suggesting that the possibility of having made eight  such errors is very unlikely.
}

{
\color{black}

\section{Radial pair distribution functions}
\label{app:gr}

In this Section we report radial pair ditribution functions, $g(r)$. We plot the O-O and the Si-Si (for the pure-silica and the 1:1 mixture).
An interesting finding concern the O-O $g(r)$ across the different compounds, see Fig.~\ref{fig:o_o_gr}.
We notice that the position of first-peak of the O-O function is similar in both the pure-water and the pure-silica compound, if one consider similar pressures.
The pure-water O-O $g(r)$ are consistent with ~\citep{cheng_phase_2021}, over the density range considered here. The position of the first peak of low-density pure-silica is consistent with the stishovite crystal structure\citep{ross1990high}, which feature O-O distances between $2.3$ and $2.6$ Angstrom, in the range 0-15 GPa.
Moreover, the first-peak height change more drastically in the pure-water (from 2 to 4) and the 1:4 mixture (from 2.2 to 3.5) case, compared to the 1:1 mixture (from 2.4 to 3) and pure-silica (from 2.5 to 3). This is consistent with the existence(absence) of a liquid-solid transition upon compression in the first(last) two materials, in the pressure range considered.

\begin{figure}[h!]
    \includegraphics[height=12cm]{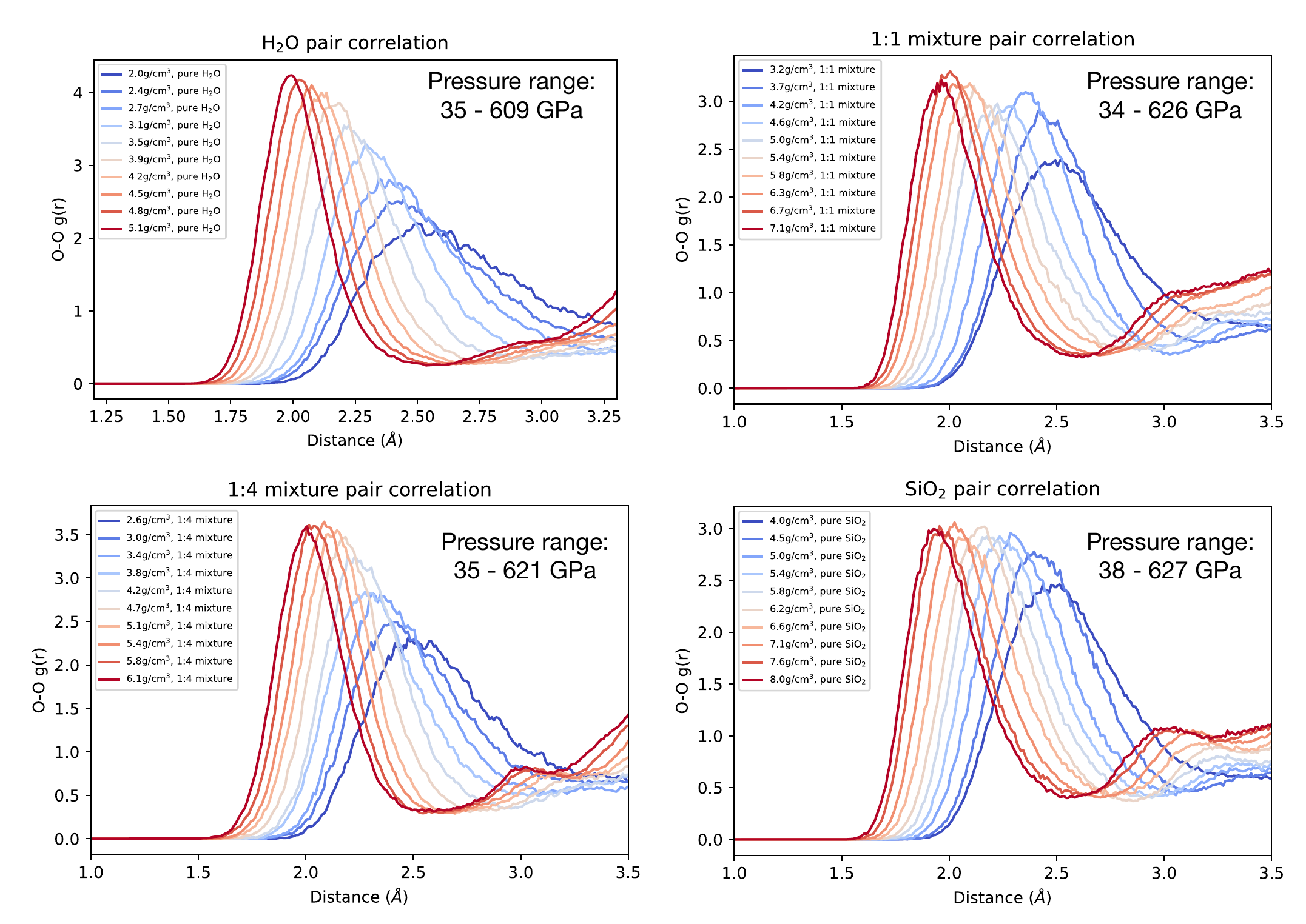}
\caption{O-O radial pair distribution functions for the pure-silica and the 1:1 mixture, at all densities. All the four compounds behave in a consistent.
The first peak position shifts towards shorter values as the density is increased, from about $2.5$ to  $2.0$ Angstrom.
}
\label{fig:o_o_gr}
\end{figure}

In Fig.~\ref{fig:si_si_gr} we also plot the $Si$-$Si$ radial pair distribution function.
From this plot, no significant differences in the two distributions can be observed. It is not possible to infer signatures of demixing. The fact that the two compression pathways, which are independent, reach compatible final g(r) (i.e. at high density) suggests that none of the two simulations has equilibrated into a metastable state.

\begin{figure}[h!]
    \includegraphics[height=6.5cm]{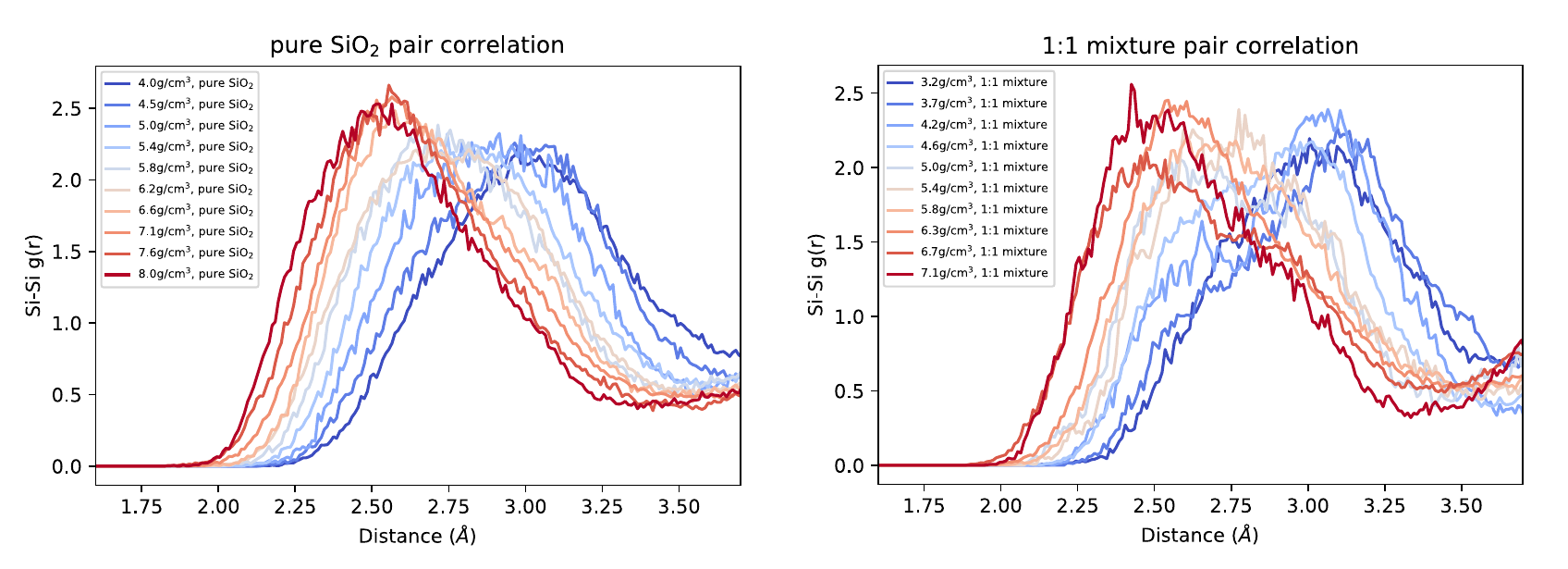}
\caption{$Si$-$Si$ radial pair distribution functions for the pure-silica and the 1:1 mixture, at all densities. The two compounds behave in a qualitatively similar way.
The first peak position shifts towards shorter values as the density is increased.
The 1:1 mixture plot appears to be noisier because there are less Si atoms in the configurations.
}
\label{fig:si_si_gr}
\end{figure}

}

\bibliographystyle{aasjournal}

\end{document}